\documentclass[lettersize,journal]{IEEEtran}
\usepackage[utf8]{inputenc}
\usepackage[T1]{fontenc}

\usepackage{enumitem}
\usepackage{amsmath, amssymb}
\usepackage{semantic}
\usepackage{nicematrix}
\usepackage[caption=false,font=normalsize,labelfont=sf,textfont=sf]{subfig}
\usepackage{arydshln}
\usepackage{algpseudocode}
\usepackage{algorithm}
\usepackage{graphicx}
\usepackage{comment}
\usepackage{booktabs}
\usepackage{pifont}
\usepackage{balance}
\usepackage{hhline}
\usepackage{diagbox}
\usepackage{multirow}
\usepackage{pict2e}
\usepackage{diagbox}
\usepackage{optidef}
\usepackage{mathrsfs}
\usepackage{tablefootnote}
\usepackage{hyperref}
\usepackage{cite}
\usepackage{makecell}
\usepackage{varwidth}
\usepackage{tikz,xcolor,hyperref}
\definecolor{lime}{HTML}{A6CE39}
\DeclareRobustCommand{\orcidicon}{
	\begin{tikzpicture}
	\draw[lime, fill=lime] (0,0) 
	circle [radius=0.16] 
	node[white] {{\fontfamily{qag}\selectfont \tiny ID}};
	\draw[white, fill=white] (-0.0625,0.095) 
	circle [radius=0.007];
	\end{tikzpicture}
	\hspace{-2mm}
}
\foreach \x in {A, ..., Z}{\expandafter\xdef\csname orcid\x\endcsname{\noexpand\href{https://orcid.org/\csname orcidauthor\x\endcsname}
			{\noexpand\orcidicon}}
}
\usepackage[compact]{titlesec}         
\titlespacing{\section}{8pt}{8pt}{8pt} 
\AtBeginDocument{
  \setlength\abovedisplayskip{2pt}
  \setlength\belowdisplayskip{2pt}} 

\newtheorem{proposition}{Proposition}

\newtheorem{definition}{Definition}

\newcommand{\Rn}{\mathbb{R}^n}
\newcommand{\Rnn}{\mathbb{R}^{n \times n}}

\newcommand{\qed}{\hfill $\square$}

\newcommand{\bac}[1]{\textcolor{black}{#1}}
\newcommand{\cac}[1]{\textcolor{cyan}{#1}}

\usepackage{soul}
\allowdisplaybreaks
\def\BibTeX{{\rm B\kern-.05em{\sc i\kern-.025em b}\kern-.08em
    T\kern-.1667em\lower.7ex\hbox{E}\kern-.125emX}}
\begin{document}
\title{
Voltage Stability of Inverter-Based Systems: Impact of Parameters and Irrelevance of Line Dynamics
}
\author{Sushobhan Chatterjee\orcidA{} and Sijia Geng\orcidB{}\\
\textit{{Dept. of Electrical and Computer Engineering}}\\
\textit{Johns Hopkins University}\\
{Baltimore, MD, USA}
\thanks{This work was supported by the U.S. National Science Foundation Global Centers Award 2330450 (EPICS) and the Ralph O'Connor Sustainable Energy Institute (ROSEI) at Johns Hopkins University. (Corresponding author: Sijia Geng, \texttt{sgeng@jhu.edu}).}%
}

\maketitle

\begin{abstract} 
This paper investigates voltage stability in inverter-based power systems concerning fold and saddle-node bifurcations. An analytical expression is derived for the sensitivity of the stability margin using the normal vector to the bifurcation hypersurface. Such information enables efficient identification of effective control parameters in mitigating voltage instability. \bac{Comprehensive analysis reveals that reactive loading setpoint and current controller's feedforward gain are the most influential parameters for enhancing voltage stability in a grid-following (GFL) inverter system, while the voltage controller's feedforward gain plays a dominant role in a grid-forming (GFM) inverter.} Notably, both theoretical and numerical results demonstrate that transmission line dynamics have no impact on fold/saddle-node bifurcations in these systems. Results in this paper provide insights for efficient analysis and control in future inverter-dominated power systems through reductions in parameter space and model complexity.
\end{abstract}

\begin{IEEEkeywords}
Voltage stability, inverter-based resources, saddle-node bifurcation, line dynamics, normal vectors, stability margin, parameter sensitivity
\end{IEEEkeywords}

\section{Introduction}
In recent years, the growing integration of inverter-based resources (IBRs), such as renewable generation and energy storage, has introduced opportunities and challenges for power system operations. These challenges, exacerbated under strained conditions and severe disturbances, may cause various issues, including transient stability \cite{milligan2015alternatives}, sub-synchronous oscillations \cite{cheng2022real}, and voltage collapse \cite{wu2024approximating}.
Voltage collapse, in particular, represents a critical instability in heavily loaded power systems, often culminating in declining voltages and potential blackouts. Voltage collapse is often triggered by the loss of a stable operating equilibrium through a fold or saddle-node bifurcation \cite{dobson1992observations}, which underscores the importance of understanding the proximity of current operating points to fold or saddle-node bifurcations. 
As a result, accurately identifying the location of the bifurcations and assessing the impacts of control parameters are essential for mitigating the risks associated with voltage collapse and ensuring the reliability of future power systems \cite{dobson1992voltage}.

In conventional power systems dominated by synchronous machines, transmission line dynamics are typically ignored and modeled using algebraic power flow equations, a simplification that is justified by the relatively slow response time of synchronous machines \cite{peponides1982singular}.
However, IBRs can operate at much faster time scales, and neglecting transmission line dynamics may impact the accuracy of stability assessments. This issue has been observed in droop-controlled microgrids \cite{mariani2014model} and has been explored in various studies, including small-signal analysis of droop gains \cite{vorobev2017high}, the impacts of dynamic lines on the oscillatory instability of grid-forming (GFM) inverters \cite{chatterjee2024effects}, and dynamic performance and small-signal stability of unified inverters\cite{geng2022unified,geng2022dynamic}. 
While dynamic models are crucial for understanding voltage collapse mechanisms, assessing loading margins to fold bifurcations primarily relies on the static components of the system. For example, a dynamic power system model $\dot{x} = f(x, \lambda)$ can be reduced to its static counterpart $f(x, \lambda) = 0$ without affecting the calculation of loading margins \cite{dobson2011irrelevance}. \bac{Building on this understanding, this paper makes two key contributions: (1) Demonstrating that line dynamics do not affect fold or saddle-node bifurcations in two widely used GFM and grid-following (GFL) inverter systems, and (2) Exploring the stability boundary and parameter sensitivity associated with voltage collapse in IBR-based systems.}
We highlight the critical role of fold or saddle-node bifurcations in voltage instability. Using the \textit{normal vector} approach\cite{chatterjee2024sensitivity}, the study derives an analytical formula to quantify how the stability margin of saddle-node bifurcation responds to parameter changes, providing valuable insights for mitigating collapse risks. Consequently, strategies can be devised to prevent the system from approaching the fold or saddle-node bifurcations.

\section{Preliminaries} \label{prelims}
This section introduces the mathematical notation and provides a concise primer on fold and saddle-node bifurcations. 

\subsection{Notations} 
Let $\Rn$ and $\mathbb{C}^n$ represent n-dimensional real and complex vector spaces, while $\Rnn$ denotes real square matrices of size n. $\mathbb{S}^n$ represents a unit hypersphere in $\mathbb{R}^n$, and $I_n$ is the identity matrix of size $n$. For any matrix $A \in \mathbb{R}^{n \times n}$, we use $\mu_i$ to indicate its $i$-th eigenvalue, and $\det(A)$ to denote its determinant. For a complex vector $x$, we use $x_i$ for its $i$-th component, $\bar{x}$ for its complex conjugate, $x^T$ and $x^H$ for its transpose and conjugate transpose, respectively, and $\Vert x \rVert$ for its Euclidean norm unless specified otherwise. For a complex number $x$, $|x|$ denotes its magnitude, while $\Re(x)$ and $\Im(x)$ indicates its real and imaginary components, respectively. Finally, $D_xf$ or $f_x$ indicates the partial derivative of function $f$ with respect to $x$.

\subsection{Fold and Saddle-Node Bifurcations} \label{sec:bifn}
Consider a system described by the differential equation,  
\begin{equation}\label{eq:sys}
    \dot{x} = f(x, \lambda),    
\end{equation}
where $f: \mathbb{R}^n \times \mathbb{R}^m \to \mathbb{R}^n$ is a smooth nonlinear function, $x \in \mathbb{R}^n$ are the state variables, and $\lambda \in \mathbb{R}^m$ are system parameters. For a specific parameter vector $\lambda_0 \in \mathbb{R}^m$, representing the nominal system parameters, let $x_0$ be an equilibrium point which is assumed to be asymptotically stable. As $\lambda$ varies, $x_0$ may change or become unstable due to bifurcations. In this analysis, $x_{*}$ and $\lambda_{*}$ denote the equilibrium and parameter values at a saddle-node bifurcation, and $f_x|_{*}$ represents the Jacobian $f_x$ evaluated at $(x_{*}, \lambda_{*})$. 

Without structural constraints, fold/saddle-node and Hopf bifurcations are generically observed for one-dimensional parameter variations in $\mathbb{R}^m$ \cite{sotomayor1973generic}. Previous works of the authors have investigated Hopf bifurcations in GFM \cite{chatterjee2024effects} and GFL \cite{chatterjee2024sensitivity} systems, which are associated with oscillatory instability. This paper focuses on fold or saddle-node bifurcations due to their pivotal role in voltage instability. To set the stage for discussion, we provide definitions for fold and saddle-node bifurcations and elaborate on their distinctions.  

\begin{definition} \label{defn:d1}
    \textit{(\cite{seydel2009practical}, Fold Bifurcation) Referring to system \eqref{eq:sys}. Assume that $f$ is a twice continuously differentiable function, and,
    \begin{enumerate}
        \item $f(x_{*},\lambda_{*}) = 0$,
        \item \textit{rank}$\{f_x(x_{*},\lambda_{*})\} = n-1$,
        \item $f_{\lambda}(x_{*},\lambda_{*}) \notin range\{f_x(x_{*},\lambda_{*})\}$, i.e., \\  $rank\big\{[f_x(x_{*},\lambda_{*})|f_{\lambda}(x_{*},\lambda_{*})]\big\} = n$,
        \item there exists parametrization $x(\sigma), \lambda(\sigma)$ with $x(\sigma_{*}) = x_{*}$, $\lambda(\sigma_{*}) = \lambda_{*}$, and $\lambda_{\sigma \sigma}(\sigma_{*}) \neq 0$.
    \end{enumerate}
    Then, the equilibrium point $(x_{*},\lambda_{*})$ is termed as the fold bifurcation point of stationary solutions.}
\end{definition}

The fold bifurcation described by Definition~\ref{defn:d1} is a generic codimension-1 bifurcation\footnote{For more context, refer to singularity theory \cite{golubitsky2012stable}.}. 
It is beneficial to exclude the rare codimension-2 scenario of a non-simple zero eigenvalue \cite[Section 7.3]{guckenheimer2013nonlinear}, i.e., the non-trivial cases of condition 2) in Definition~\ref{defn:d1}. By further enforcing the simplicity of the zero eigenvalue, a generic saddle-node bifurcation is identified, whose definition can be stated as follows.
\begin{definition} \label{defn:d2}
    \textit{(\cite{dobson2011irrelevance}, Saddle-Node Bifurcation) Referring to system \eqref{eq:sys}. Assume that $f$ is a twice continuously differentiable function, and,
    \begin{enumerate}
        \item[a)] Definition~\ref{defn:d1} is satisfied.
        \item[b)] $f_x|_{*}$ has a simple zero eigenvalue.
    \end{enumerate}
    Then, the equilibrium point $(x_{*},\lambda_{*})$ is termed as a generic saddle-node bifurcation point of stationary solutions.}
\end{definition}

Voltage instability and collapse are often tied to singularities caused by these bifurcations \cite{54571}.

\section{Irrelevance of Line Dynamics}\label{sec:irre}
In this section, we demonstrate that line dynamics are irrelevant in voltage instability for a class of IBR systems. In other words, ignoring the dynamics in line models does not affect the margins to fold bifurcations or their sensitivities. This insight simplifies modeling efforts and offers practical benefits. The following proposition formalizes the conditions under which fold bifurcations remain unaffected by the reduction of dynamic elements.  

\begin{proposition} \label{prop:p1} \cite[Lemma 5]{dobson2011irrelevance} Let $f(x,\lambda) = \begin{pmatrix}
    f^1(a,b,\lambda)\\
    f^2(a,b,\lambda)
\end{pmatrix}$ be a smooth function with component functions $f^1 : \mathbb{R}^{\tilde{n}} \times \mathbb{R}^{n-\tilde{n}} \times \mathbb{R}^m \rightarrow \mathbb{R}^{\tilde{n}}$ and $f^2 : \mathbb{R}^{\tilde{n}} \times \mathbb{R}^{n-\tilde{n}} \times \mathbb{R}^m \rightarrow \mathbb{R}^{{n-\tilde{n}}}$ respectively. Suppose that rank($f_b^2|_{*}) = n - \tilde{n}$ so that $f_b^2|_{*}$ is invertible. Define $\bar{f} : \mathbb{R}^{\tilde{n}} \times \mathbb{R}^m \rightarrow \mathbb{R}^{\tilde{n}}$, such that $\dot{a} = f^1(a,\phi(a,\lambda),\lambda):= \bar{f}(a,\lambda) $, where $\phi : U \rightarrow \mathbb{R}^{n-\tilde{n}}$ is the unique smooth function  satisfying $f^2(a,\phi(a,\lambda),\lambda) = 0$ and $\phi(a_{*},\lambda_{*}) = b_{*}$ for some neighbourhood $U$ of $(a_{*},\lambda_{*})$. Write $z := (x,\lambda)$ and $\bar{z} := (a,\lambda)$. Then,
\begin{enumerate}
    \item $\text{rank}(f_z) = n \Leftrightarrow \text{rank}(\bar{f}_{\bar{z}}) = \tilde{n}$,
    \item $f$ has a fold bifurcation at $z_{*} := (a_{*},b_{*},\lambda_{*}) \Leftrightarrow \bar{f}$ has a fold bifurcation at $\bar{z}_{*} := (a_{*},\lambda_{*})$. \qed
\end{enumerate} 
\end{proposition}  

\begin{figure}[ht!]  
\hspace{0.3ex} \centerline{\includegraphics[scale=0.87]{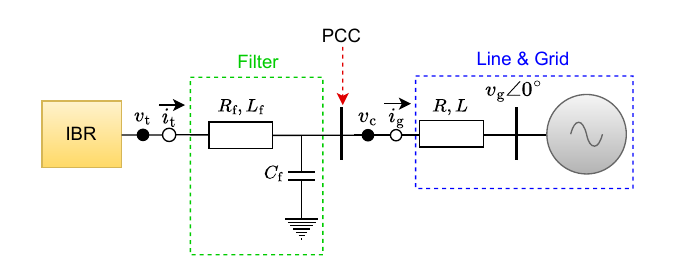}}  
\caption{General architecture of a single-inverter-infinite-bus system.}  
\label{fig:genl}  
\end{figure}

Building on this insight, we proceed to show that ignoring line dynamics does not affect fold bifurcations for a class of IBRs in a configuration where a single inverter is connected to the grid (infinite bus) via a transmission line (series resistance and inductance), as illustrated in Fig. \ref{fig:genl}. \bac{Here, $v_\text{t}$ and $i_\text{t}$ represent the IBR's terminal voltage and current, while $v_\text{c}$ and $i_\text{g}$ denote the corresponding values at the point-of-common-coupling (PCC). With a slight abuse of notation, $v_\text{g}$ is the grid voltage. $R_\text{f},L_\text{f},C_\text{f}$ are the output filter's resistance, inductance, and capacitance, and $R,L$ are the line's resistance and inductance.}
\vspace{-4ex}

Let the overall system dynamics be represented as,  
\begin{align}  
\dot{x} = h(x, \lambda), \label{eq:h}
\end{align}   
where $x$ can be partitioned into two components, with $x_1 \in \mathbb{R}^{n-q}$ representing the dynamic states of the inverter and $x_2 \in \mathbb{R}^q$ the dynamic states of the line. We can write,
\begin{align}
\dot{x}_1 &= h_1(x_1, x_2, \lambda), \\
\dot{x}_2 &= h_2(x_1, x_2, \lambda).
\end{align}
\bac{For the system under investigation, $x_2 = (i_\text{g,D}, i_\text{g,Q}) \in \mathbb{R}^2$, that is, the projection of $i_\text{g}$ onto the global D- and Q-axis (refer to \cite{geng2022unified,chatterjee2024effects} for more on reference frames). The dynamic equations for the line are given (in per unit) as,}
\vspace{-0.3ex}
\begin{align}  
\dot{i_\text{g,D}} - \frac{\omega_\text{b}}{L}(v_\text{c,D} - v_\text{g,D}) + \frac{R}{L}\omega_\text{b}i_\text{g,D} - \omega_\text{DQ}\omega_\text{b}i_\text{g,Q} &= 0, \label{eqn:sm12} \\  
\dot{i_\text{g,Q}} - \frac{\omega_\text{b}}{L}(v_\text{c,Q} - v_\text{g,Q}) + \frac{R}{L}\omega_\text{b}i_\text{g,Q} + \omega_\text{DQ}\omega_\text{b}i_\text{g,D} &= 0. \label{eqn:sm13}  
\end{align}  
\bac{Here, $v_\text{g,D}$ and $v_\text{g,Q}$ are the D- and Q-axis grid voltages, $v_\text{c,D}$ and $v_\text{c,Q}$ are the D- and Q-axis PCC voltages, $\omega_\text{b}$ is the base angular frequency in rad/s, and $\omega_\text{DQ}$ is the per-unit frequency of the global DQ-frame. The electrical variables in the DQ-frame become constant at steady state when $\omega_\text{DQ}$ aligns with the steady-state frequency $\omega_\text{ss}$ (which equals $\omega_0$ in the presence of an infinite bus). As a result, the dynamic line model in equations \eqref{eqn:sm12} and \eqref{eqn:sm13} reduces to a set of algebraic equations at steady state, } 
\begin{equation}
    h_2(x_1, x_2, \lambda) = 0.\label{eq:ss}
\end{equation}

To analyze the effects of line dynamics, we first demonstrate the non-singularity of the Jacobian of the line dynamics. Computing the Jacobian yields,

\begin{align*}
D_{x_2}h_2 =  
\begin{bmatrix}  
-\frac{R}{L}\omega_\text{b} & \omega_\text{DQ}\omega_\text{b} \\  
-\omega_\text{DQ}\omega_\text{b} & -\frac{R}{L}\omega_\text{b}  
\end{bmatrix},  
\end{align*}  
with \(\det(D_{x_2}h_2) = \left(\omega_\text{DQ}^2 + \frac{R^2}{L^2}\right)\omega_\text{b}^2 \neq 0\). Therefore, the Jacobian is always non-singular, regardless of the system state. By the Implicit Function Theorem, a unique local solution exists for the system \eqref{eq:ss},  
allowing us to express \(x_2\) as a smooth function of \(x_1\) and \(\lambda\), 
\[
x_2 = g(x_1, \lambda).  
\]  

This relation can be explicitly written as,  
\begin{align}  
i_\text{g,D} &= \frac{1}{L(\omega_\text{DQ}^2 + \frac{R^2}{L^2})}\left[\frac{R}{L}(v_\text{c,D} \!-\! v_\text{g,D}) + \omega_\text{DQ}(v_\text{c,Q} \!-\! v_\text{g,Q})\right] \nonumber \\  
&:= g_1(x_1, \lambda), \\  
i_\text{g,Q} &= \frac{1}{L(\omega_\text{DQ}^2 + \frac{R^2}{L^2})}\left[\frac{R}{L}(v_\text{c,Q} \!-\! v_\text{g,Q}) - \omega_\text{DQ}(v_\text{c,D} \!-\! v_\text{g,D})\right] \nonumber \\  
&:= g_2(x_1, \lambda).  
\end{align}

Hence, the general $n$-th order single-inverter-infinite-bus system \eqref{eq:h} reduces to an $(n-2)$-th order system with, 
\begin{equation}
    \dot{x}_1 = h_1(x_1, g(x_1, \lambda), \lambda):=\bar{h}(x_1, \lambda). 
\end{equation}
Consequently, by Proposition \ref{prop:p1}, the fold bifurcation of $h$ at $({x_1}_{*}, {x_2}_{*}, \lambda_{*})$ is equivalent to the fold bifurcation of $\bar{h}$ at $({x_1}_{*}, \lambda_{*})$. This establishes that fold bifurcations in the IBR system with dynamic line model are equivalent to those in the system with static line model.

\section{Stability Margin to Bifurcations}
To prevent voltage collapse, it is essential to characterize the stability margin to bifurcations and design measures to avoid approaching the stability boundary. The hypersurfaces corresponding to saddle-node bifurcations, denoted as $\Sigma^\text{SN} \subseteq \mathbb{R}^m$, represent the set of parameters $\lambda_{*}$ where saddle-node bifurcations occur at $(x_{*}, \lambda_{*})$. For a given parameter value $\lambda_0$, important questions are how close $\lambda_0$ is to the set $\Sigma^\text{SN}$, and what are the most effective control parameters to enlarge this margin.

\subsection{Normal Vectors to Bifurcation Hypersurfaces}
Consider $f_x|_{*}$ having a unique \textit{simple} zero eigenvalue at $(x_{*}, \lambda_{*})$. Furthermore, assume that $\lambda_{*} \in \Sigma^\text{SN}$ satisfies the non-degeneracy and transversality conditions \cite[pp. 198]{schaeffer1985singularities}, given respectively by,
    \begin{align}\label{eqn:e2}
        &\ w_{*}f_{xx}|_{*} \neq 0,
        \\&\
        w_{*}^Hf_{\lambda}|_{*} \neq 0,\label{eq:tran}
    \end{align}
where $v_{*}$ and $w_{*}$ are the right and left eigenvectors of $f_x|_{*}$ corresponding to the zero eigenvalue, i.e., $f_x|_{*}v_{*} = 0$ and $w^H_{*}f_x|_{*} = 0$. 

\bac{Given any such $\lambda_{*} \in \Sigma^\text{SN}$, there exists an open neighbourhood $V \ni \lambda_{*}$ such that the intersection $\Sigma^\text{SN} \cap V$ forms a smooth hypersurface \cite[pp. 217]{chow2012methods}. Furthermore, in the vicinity of the bifurcation point $z_{*} = (x_{*},\lambda_{*})$, there exists a unique smooth function $u$ and a scalar constant $\epsilon > 0$, such that $x = u(\lambda)$ and $f(u(\lambda),\lambda) = 0$ for all $\lVert z - z_{*} \rVert \leq \epsilon$. It follows that $\Sigma^\text{SN}$ has a normal vector $\mathcal{N}(\lambda_{*})$ at $\lambda_{*} \in \Sigma^\text{SN}$ and the associated Gauss map $\mathcal{N} : \Sigma^\text{SN} \rightarrow \mathbb{S}^m$ is smooth.} Moreover, using the transversality condition \eqref{eq:tran}, we can derive an expression for the normal vector to the saddle-node bifurcation. 

Let $d\lambda_{*}$ be an infinitesimally small arbitrary shift of $\lambda_{*}$ in $\Sigma^\text{SN}$, i.e., a differential one-form in the co-tangent bundle $T(\Sigma^\text{SN})_{\lambda_{*}} \ni d\lambda_{*}$. Then, 
\begin{equation} \label{eqn:ea1}
    \begin{aligned}
        &\ d\{f(x,\lambda)\}|_{*} = f_x|_{*}dx_{*} + f_{\lambda}|_{*}d\lambda_{*} = 0,
    \end{aligned}
\end{equation}
and,
\begin{equation}
\label{eqn:eaa1}
w_{*}^H f_{\lambda}|_{*}d\lambda_{*} = - w_{*}^H f_x|_{*}dx_{*} = 0.
\end{equation}
Since $w_{*}^H f_{\lambda}|_{*} \neq 0$ per the transversality condition \eqref{eq:tran}, and $d\lambda_{*}$ is arbitrarily chosen, so \eqref{eqn:eaa1} is satisfied only when $w_{*}^Hf_{\lambda}|_{*}$ is orthogonal to $d\lambda_{*}$, which implies that $w_{*}^H f_{\lambda}|_{*}$ is a normal vector to $\Sigma^\text{SN}$ at $\lambda_{*}$.
Therefore, the normal vector formula is given by,
\begin{equation} \label{eqn:e3}
    \mathcal{N}(\lambda_{*}) = \alpha w_{*}^Hf_{\lambda}|_{*},
\end{equation}
where $\lambda_{*} \in \Sigma^\text{SN}$ and $\alpha$ is a nonzero real scale factor chosen so that $|\mathcal{N}(\lambda_{*})| = 1$, and the sign of $\alpha$ is such that changing $\lambda$ in the direction of $\mathcal{N}(\lambda_{*})$ leads to disappearance of $x_0$.

\subsection{Parameter Sensitivity of Stability Margin}
\bac{The stability margin, defined as the distance to the bifurcation hypersurface $\Sigma^\text{SN}$ from a nominal parameter $\lambda_0$ in the direction $k$, is written as $\Delta(\lambda_0) = \lVert \lambda_{*} - \lambda_0 \rVert$.}  When $\Delta$ is small, it is beneficial to adjust $\lambda_0$ to increase $\Delta$. The optimal adjustment direction (in parameter space) is determined by the sensitivity $\Delta_{\lambda_0} = \frac{\partial \Delta(\lambda_0)}{\partial \lambda_0}$, which identifies the most effective parameter changes that enlarge the margin.

\begin{figure}[ht!]
    \vspace{-4ex} \hspace{-2ex}\centerline{\includegraphics[scale=1.8]{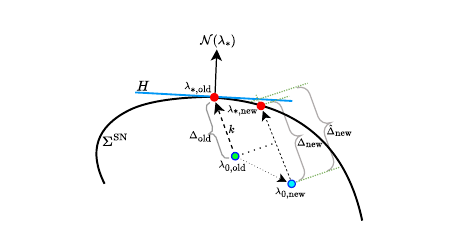}} 
    \vspace{-4ex}
    \caption{Geometry of the bifurcation hypersurface, normal vector, and the stability margin along direction $k$. \bac{$H$ is the tangent hyperplane at $\lambda_{*,\text{old}}$. With the cause of bifurcation remaining constant (i.e., moving along the direction $k$), the true stability margin increases from $\Delta_\text{old}$ to $\Delta_\text{new}$ as the operating point shifts from $\lambda_{0,\text{old}}$ to $\lambda_{0,\text{new}}$. $\hat{\Delta}_\text{new}$ is an estimation of true margin $\Delta_\text{new}$.}}
    \label{fig:norm_vec} 
\end{figure}

\bac{Figure \ref{fig:norm_vec} illustrates the bifurcation, as the parameters vary in the specific direction $k \in \mathbb{R}^m$.} 
\bac{Building on \cite{chatterjee2024effects}, we apply the condition for the first bifurcation, $\Re\{\mu(\lambda_{*})\} = 0$, and derive the following expression,} 
\begin{align*}
    \mathcal{N}(\lambda_{*})^T\left(I_m + k\Delta_{\lambda_0}^T\right) = 0.
\end{align*}
This in-turn leads to the following sensitivity expression,
\begin{align*}
    \Delta_{\lambda_0} = -\left[k^T\mathcal{N}(\lambda_{*})\right]^{-1} \mathcal{N}(\lambda_{*}).
\end{align*}
Parameters $\lambda$ are divided into uncontrollable ($\lambda^\text{uc}$) and controllable ($\lambda^\text{c}$) components. While $\lambda^\text{uc}$ may cause bifurcations and cannot be adjusted, $\lambda^\text{c}$ can be modified to improve robustness against variations in $\lambda^\text{uc}$. For this, the normal vector is partitioned as $\mathcal{N} = (\mathcal{N}^\text{uc}, \mathcal{N}^\text{c})$. For parameter variations along $k$ with only uncontrollable components (i.e., $k^\text{c} = 0$), the sensitivity with respect to $\lambda^\text{c}$ is,  
\begin{equation} \label{eqn:s1}
    \Delta_{\lambda^\text{c}_0} = -\big[(k^\text{uc})^T\mathcal{N}^\text{uc}(\lambda_{*})\big]^{-1} \mathcal{N}^\text{c}(\lambda_{*}).
\end{equation}
This sensitivity guides adjustments to $\lambda^\text{c}$ for maximizing the stability margin of $\lambda^\text{uc}$.

\section{Numerical Results} \label{res}
\begin{figure}[ht!]
    \hspace{-2ex}\centerline{\includegraphics[scale=0.65]{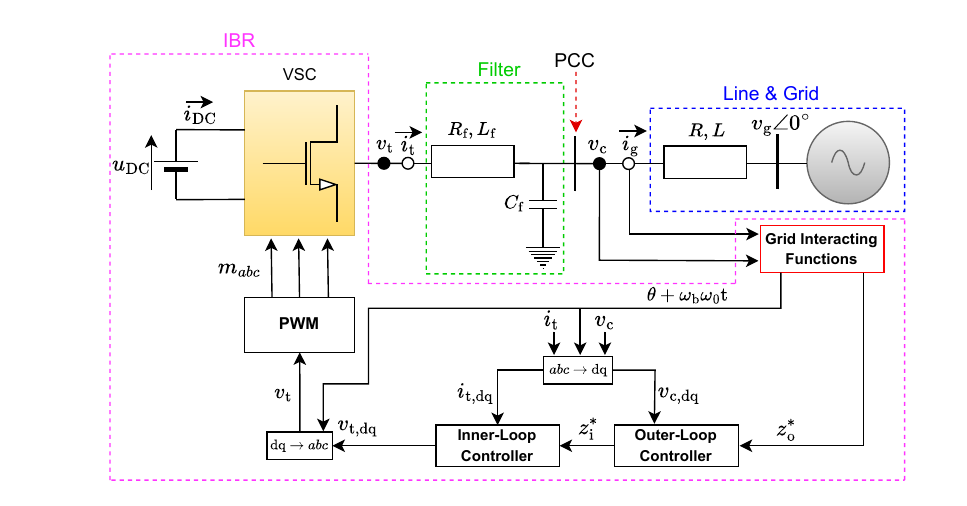}}
    \caption{Schematic of a generic IBR connecting to the grid.}
    \label{fig:genr} 
\end{figure}

This section presents numerical results that examine the impacts of various parameters on saddle-node bifurcations\footnote{For the case studies, results show a simple zero eigenvalue at fold bifurcations, confirming their saddle-node nature. Thus, fold and saddle-node bifurcations are used interchangeably.} for two common GFM and GFL inverters, whose models are illustrated in the unified schematic as depicted in Fig. \ref{fig:genr}. We identify key control parameters that can prevent voltage collapse instability and verify the irrelevance of line dynamics for voltage stability analysis. 
\vspace{-4ex}
\vspace{-1ex}

\subsection{GFL Inverters} 
Most IBRs currently operating in power grids are designed as GFL systems. These units typically synchronize with the grid using phase-locked loops (PLLs) and are designed to deliver specific amounts of active and reactive power. The cascading control scheme includes an outer loop for power control and an inner loop for current control. \bac{The detailed system model is provided in the Appendix.}

\subsubsection{System Setup and Ground Truth}
The controllable and uncontrollable parameters in the GFL system are summarized as,
\begin{equation*}
    \begin{aligned}
        & \lambda^\text{uc} := \big(R_\text{f},L_\text{f},C_\text{f},\omega_{\{0,\text{b}\}},\omega_{\{pc,qc\}},v_\text{g,\{D,Q\}},L,R\big) \in \mathbb{R}^{11},
        \\
        & \lambda^\text{c} := \big(K^\text{\{P,I\}}_\text{PLL},K^\text{\{P,I,F\}}_\text{CC},K^\text{\{P,I\}}_\text{APC},K^\text{\{P,I\}}_\text{RPC},p^{*},q^{*}\big) \in \mathbb{R}^{11}.
    \end{aligned}
\end{equation*} 
Table \ref{tab:tgfl} lists the nominal parameters, with the notations defined in the Appendix. \bac{All parameters are given in per units, except for $\omega_{\{pc,qc,\text{b}\}}$, which are in rad/s. Throughout the analysis, we set $v_\text{g,D} = 1$ p.u. and $v_\text{g,Q} = 0$ p.u.}
\begin{table}[ht!]
\vspace{-3.5ex}
    \centering
  \setlength\tabcolsep{3.2pt}
  \setlength\extrarowheight{2pt}
  \caption{Nominal parameters for GFL inverter.}
  \vspace{-1ex}
    \begin{tabular}{cccccccccc}
      \Xhline{2\arrayrulewidth} 
       $R_\text{f}$ & $L_\text{f}$ & $C_\text{f}$ & $\omega_0$ &  $\omega_{pc}$ & $\omega_{qc}$ & $K^\text{P}_\text{PLL}$ & $K^\text{I}_\text{PLL}$ & $K^\text{P}_\text{APC}$ & $K^\text{I}_\text{APC}$ \vspace{0.8ex}\\
      \hline
      0.0072 & 0.05 & 0.25 & 1 & 332.8 & 732.8 & 2.65 & 6.5 & 1 & 3\\
      \hhline{|==========|}
       $K^\text{P}_\text{RPC}$ & $K^\text{I}_\text{RPC}$ & $K^\text{P}_\text{CC}$ & $K^\text{I}_\text{CC}$ & $K^\text{F}_\text{CC}$ & $p^{*}$ & $q^{*}$ & $\omega_\text{b}$ & $L$ & $R$ \vspace{0.8ex}\\ 
       \hline
      -3 & -4 & 0.3 & 3 & 0 & 1 & 0.5 & 120$\pi$ & 0.07 & 0.02 \\ 
      \Xhline{2\arrayrulewidth} 
    \end{tabular}
    \label{tab:tgfl}
    \vspace{-1.5ex}
\end{table}

The parameter values that (individually) trigger saddle-node bifurcations are listed in Table \ref{tab:GFL_param}. A key observation is that all the other parameters do not lead to a saddle-node bifurcation except for the \bac{active and reactive power} setpoints. 
It also compares results for static and dynamic transmission line models and numerically verifies that stability margins are unaffected by the neglect of transmission line dynamics. This aligns with the theoretical result derived in Section~\ref{sec:irre}.  
\begin{table*}[htbp]
  \setlength\tabcolsep{0pt}
  \setlength\extrarowheight{2pt}
  \caption{\bac{Instability-inducing bifurcation parameters} \bac{(in per units)} for GFL.}
  \makebox[\textwidth][c]{
    \begin{tabular*}{\textwidth}{@{\extracolsep{\fill}}*{6}{c}}
      \Xhline{2\arrayrulewidth} 
      \multirow{2}{*}{Parameter \bac{($I$)}} & \multirow{2}{*}{Nominal value \bac{($\lambda_0$)}} & \multicolumn{2}{c}{Value at saddle-node bifurcation \bac{($\lambda_{*}$)}} & \multicolumn{2}{c}{True margin \bac{($\Delta$)}} \\
      \cline{3-4} \cline{5-6}
       & & \multicolumn{1}{c}{S-line} & \multicolumn{1}{c}{D-line} & \multicolumn{1}{c}{S-line} & \multicolumn{1}{c}{D-line} \\
      \hline
      $p^{*}$ & 1  & 10.10247  & 10.10247  & 9.10247  & 9.10247  \\
      $q^{*}$ & 0.5  & 64.678492  & 64.678321  & 64.178492  & 64.178321  \\
     \Xhline{2\arrayrulewidth}
     \end{tabular*}
     }
    \label{tab:GFL_param}
\end{table*}

\subsubsection{Sensitivity Analysis of Control Parameters Using Normal Vectors}
The onset of saddle-node bifurcation can be prevented through control tuning using the sensitivity information \eqref{eqn:s1} provided by the normal vector. The estimated stability margin $\hat{\Delta}^I_\text{new}$ is given by,  
\bac{\begin{equation} \label{ee}
    \hat{\Delta}^I_\text{new} = \Delta^I_\text{old} + \Delta^{I|C}_{\lambda_0}\big(\lambda^C_{0,\text{new}} - \lambda^C_{0,\text{old}}\big),
\end{equation}}
where $I$ and $C$ represent the indices of the instability-inducing and control parameters, respectively. 

The true margin is expressed as,  
\bac{\begin{equation} \label{er}
    \Delta^I_\text{new} = \lambda^I_{*,\text{new}} - \lambda^I_{0,\text{new}},
\end{equation}}
which is calculated numerically by finding $\lambda^I_{*,\text{new}}$. Compared to \eqref{er}, the estimation based on the normal vector is much more computationally efficient. Tables \ref{tab:GFL_infl_S} and \ref{tab:GFL_infl_D} summarize quantitatively the impacts of all system parameters on mitigating instabilities caused by the instability-inducing parameters $I$ as identified in Table~\ref{tab:GFL_param}, for static and dynamic lines, respectively.
\begin{table}[ht!]
\vspace{-3ex}
    \begin{center}
  \setlength\tabcolsep{1.3pt}
  \setlength\extrarowheight{2pt}
   \caption{\vspace{0.8ex} \hspace{2ex} Stability Margin Sensitivity $\big(\Delta^{I|C}_{\lambda_0}\big)$ of GFL with static line.}
   \vspace{-2ex}
    \begin{tabular}{|r||*{11}{c|}}
      \Xhline{2\arrayrulewidth} \vspace{-0.5ex} 
      \diagbox[outerleftsep=-3.8pt,outerrightsep=-2.8pt]{\scriptsize{$I\vspace{0.5ex}$}}{\scriptsize{$C\hspace{0.25ex}$}} & \scriptsize{$K^\text{P}_\text{PLL}$} & \scriptsize{$K^\text{I}_\text{PLL}$} & \scriptsize{$K^\text{P}_\text{CC}$} & \scriptsize{$K^\text{I}_\text{CC}$} & \scriptsize{$K^\text{F}_\text{CC}$} & \scriptsize{$K^\text{P}_\text{APC}$} &  \scriptsize{$K^\text{I}_\text{APC}$} & \scriptsize{$K^\text{P}_\text{RPC}$} & \scriptsize{$K^\text{I}_\text{RPC}$} & \scriptsize{$p^{*}$} & \scriptsize{$q^{*}$}\\
      \hline
      \scriptsize{$p^{*}$} & 0 & 0 & 0 & -0.003 & -0.01 & 0 & 0.403 & 0 & 0.011 & - & \cac{1.317}\\
      \hline
      \scriptsize{$q^{*}$} & 0 & 0 & 0 & 5.071 & \cac{15.46} & 0 & 0.481 & 0 & 2.96 & 2.606 & - \\
      \Xhline{2\arrayrulewidth} 
    \end{tabular}
    \label{tab:GFL_infl_S}
    \end{center}
\end{table} 

Notably, the reactive power setpoint $q^{*}$ turns out to be the most effective parameter in mitigating saddle-node bifurcations resulting from changes in the active power setpoint (for both static and dynamic lines). This reflects the understanding that voltage instabilities, caused by operating GFL inverters under high active loading beyond the maximum power transfer limit, can be effectively addressed through controlled reactive power support achieved through tuning $q^{*}$. \bac{Similarly, $K^\text{F}_\text{CC}$ proves to be the most effective parameter in preventing voltage collapse of the GFL system when driven by a high reactive power setpoint.}
\begin{table}[ht!]
\vspace{-2.5ex}
    \begin{center}
  \setlength\tabcolsep{1.3pt}
  \setlength\extrarowheight{2pt}
   \caption{\vspace{0.8ex} \hspace{0.4ex} Stability Margin Sensitivity $\big(\Delta^{I|C}_{\lambda_0}\big)$ of GFL with dynamic line.}
   \vspace{-2ex}
    \begin{tabular}{|r||*{11}{c|}}
      \Xhline{2\arrayrulewidth} \vspace{-0.5ex}
      \diagbox[outerleftsep=-3.8pt,outerrightsep=-2.8pt]{\scriptsize{$I\vspace{0.5ex}$}}{\scriptsize{$C\hspace{0.25ex}$}} & \scriptsize{$K^\text{P}_\text{PLL}$} & \scriptsize{$K^\text{I}_\text{PLL}$} & \scriptsize{$K^\text{P}_\text{CC}$} & \scriptsize{$K^\text{I}_\text{CC}$} & \scriptsize{$K^\text{F}_\text{CC}$} & \scriptsize{$K^\text{P}_\text{APC}$} &  \scriptsize{$K^\text{I}_\text{APC}$} & \scriptsize{$K^\text{P}_\text{RPC}$} & \scriptsize{$K^\text{I}_\text{RPC}$} & \scriptsize{$p^{*}$} & \scriptsize{$q^{*}$}\\
      \hline
      \scriptsize{$p^{*}$} & 0 & 0 & 0 & -0.003 & -0.01 & 0 & 0.404 & 0 & 0.011 & - & \cac{1.316}\\
      \hline
      \scriptsize{$q^{*}$} & 0 & 0 & 0 & 5.071 & \cac{15.41} & 0 & 0.481 & 0 & 2.78 & 2.610 & -\\
      \Xhline{2\arrayrulewidth} 
    \end{tabular}
    \label{tab:GFL_infl_D}
    \end{center}
    \vspace{-2.5ex}
\end{table}

\subsubsection{Example on Stability Margin Estimation}
Consider the active power setpoint $p^{*}$ as the destabilizing parameter and $q^{*}$ as the control parameter to counteract its effects. Tables \ref{tab:GFL_stat} and \ref{tab:GFL_dyn} give a comparison between the true margin and the estimated margin derived from the normal vector method for systems with static and dynamic transmission lines, respectively. The initial row displays the nominal values of $q^{*}$ and the corresponding true margin of $p^{*}$. The subsequent rows illustrate the increments in $q^{*}$ and the associated changes in the stability margin of $p^{*}$, for both the estimated and true values. The results demonstrate that the normal vector method provides accurate estimations of sensitivity margins. Its accuracy decreases slightly as the parameter values deviate further from their nominal settings, which is consistent with the fact that the prediction formula \eqref{ee} relies on a linear approximation. Furthermore, the margins for systems with and without line dynamics are the same except for very slight deviations due to numerical inaccuracy. 
\begin{table}[ht!]
\vspace{-2ex}
\begin{center}
\caption{Effects of parameter variation \bac{(in per units)} on stability margin for GFL: Static line.}
\vspace{-1ex}
\begin{tabular}{cccc}
\Xhline{2\arrayrulewidth}
\multirow{2}{*}{$q^{*}\ \bac{\uparrow}$} & Estimated Margin & True Margin & \multirow{2}{*}{Error} \\
& in $p^{*}$ $\bac{\uparrow}$ & in $p^{*}$ & \\
\hline
        0.5 (Nominal) & - & 9.10247 & - \\
        0.7 & 9.36577 & 9.355 & 1.08e-02 \\
         0.9 & 9.62907 & 9.6 & 2.91e-02 \\
         1.2 & 10.02402 & 9.96 & 6.4e-02 \\
          \Xhline{2\arrayrulewidth}
\end{tabular}
\label{tab:GFL_stat}
\end{center}
\end{table}

\begin{table}[ht!]
\vspace{-5ex}
\begin{center}
\caption{Effects of parameter variation \bac{(in per units)} on stability margin for GFL: Dynamic line. }
\vspace{-1ex}
\begin{tabular}{cccc}
\Xhline{2\arrayrulewidth}
\multirow{2}{*}{$q^{*}\ \bac{\uparrow}$} & Estimated Margin & True Margin & \multirow{2}{*}{Error} \\
& in $p^{*}$ $\bac{\uparrow}$ & in $p^{*}$ & \\
\hline
        0.5 (Nominal) & - & 9.10247 & - \\
        0.7 & 9.36563 & 9.353 & 1.26e-02 \\
         0.9 & 9.62879 & 9.598 & 3.08e-02 \\
         1.2 & 10.02353 & 9.959 & 6.45e-02 \\
          \Xhline{2\arrayrulewidth}
\end{tabular}
\label{tab:GFL_dyn}
\end{center}
\vspace{-3ex}
\end{table}

\subsection{GFM Inverters}
The GFM inverter functions as a controlled voltage source. The detailed mathematical model describing the droop-based GFM system is given in \cite{chatterjee2024effects}.

\subsubsection{System Setup and Ground Truth}
Similar to the GFL case, the goal is to identify the most effective control parameters to mitigate the effects of uncontrollable parameters in inducing saddle-node bifurcation. The system's parameters are categorized as follows,  
\begin{equation*}
    \begin{aligned}
        \lambda^\text{uc} & := \big(R_\text{f},L_\text{f},C_\text{f},\omega_\text{b},\omega_{\{pc,qc\}},v_\text{g,\{D,Q\}},L,R\big) \in \mathbb{R}^{10},
        \\
        \lambda^\text{c} & := \big(K^\text{\{P,I,F\}}_\text{VC},K^\text{\{P,I,F\}}_\text{CC},K_\text{\{P,Q\}},\omega_0,V_0,p^{*},q^{*}\big) \in \mathbb{R}^{12}.
    \end{aligned}
\end{equation*} 
Table \ref{tab:tgfm} provides the nominal GFM parameters adapted from \cite{chatterjee2024effects}, and the notations are explained therein.  \bac{All parameters are expressed in per units except for $\omega_{\{pc,qc,\text{b}\}}$, which are in rad/s. We assume $v_\text{g,D} = 1$ p.u. and $v_\text{g,Q} = 0$ p.u.} 
\begin{table}[ht!]
\vspace{-2ex}
    \centering
  \setlength\tabcolsep{4.2pt}
  \setlength\extrarowheight{2pt}
  \caption{Nominal parameters for GFM inverter.}
  \vspace{-1ex}
    \begin{tabular}{cccccccccc}
      \Xhline{2\arrayrulewidth} 
       $R_\text{f}$ & $L_\text{f}$ & $C_\text{f}$ & $\omega_0$ & $V_0$ &  $\omega_{pc}$ & $\omega_{qc}$ & $K^\text{P}_\text{VC}$ & $K^\text{I}_\text{VC}$ & $K^\text{F}_\text{VC}$ \vspace{0.8ex}\\
      \hline
      0.0072 & 0.05 & 0.25 & 1 & 1 & 332.8 & 732.8 & 1 & 2.5 & 1 \\
      \hhline{|==========|}
       $K^\text{P}_\text{CC}$ & $K^\text{I}_\text{CC}$ & $K^\text{F}_\text{CC}$ & $K_\text{P}$ & $K_\text{Q}$ & $p^{*}$ & $q^{*}$ & $\omega_\text{b}$ & $L$ & $R$ \vspace{0.8ex}\\ 
       \hline
      2.5 & 2.5 & 0 & 0.5 & 0.01 & 1 & 0.5 & 120$\pi$ & 0.8 & 0.2 \\ 
      \Xhline{2\arrayrulewidth} 
    \end{tabular}
    \label{tab:tgfm}
\end{table}

Table \ref{tab:GFM_param} lists parameters that (individually) trigger saddle-node bifurcations, excluding those with no impact for brevity. It also compares results for static and dynamic transmission line models. Similar to the GFL case, stability margins remain unaffected by neglecting transmission line dynamics, which is consistent with the theoretical result.
\begin{table*}[htbp]
  \setlength\tabcolsep{0pt}
  \setlength\extrarowheight{2pt}
  \caption{\bac{Instability-inducing bifurcation parameters} \bac{(in per units)} for GFM.}
  \makebox[\textwidth][c]{
    \begin{tabular*}{\textwidth}{@{\extracolsep{\fill}}*{6}{c}}
      \Xhline{2\arrayrulewidth} 
      \multirow{2}{*}{Parameter \bac{($I$)}} & \multirow{2}{*}{Nominal value \bac{($\lambda_0$)}} & \multicolumn{2}{c}{Value at saddle-node bifurcation \bac{($\lambda_{*}$)}} & \multicolumn{2}{c}{True margin \bac{($\Delta$)}} \\
      \cline{3-4} \cline{5-6}
       & & \multicolumn{1}{c}{S-line} & \multicolumn{1}{c}{D-line} & \multicolumn{1}{c}{S-line} & \multicolumn{1}{c}{D-line} \\
      \hline
      $V_0$ & 1 & 0.705092 & 0.705092 & 0.294908 & 0.294908\\
      $p^{*}$ & 1 & 1.495078 & 1.495078 & 0.495078 & 0.495078\\
      $L$ & 0.8  & 1.149634 & 1.149634 & 0.349634 & 0.349634 \\
      $R$ & 0.2  & 1.133009 & 1.133009 & 0.933009 & 0.933009\\
      \Xhline{2\arrayrulewidth}
     \end{tabular*}
     }
    \label{tab:GFM_param}
\end{table*}

\subsubsection{Sensitivity Analysis of Control Parameters Using Normal Vectors}
Tables \ref{tab:GFM_infl_S} and \ref{tab:GFM_infl_D} summarize quantitatively the impacts of all system parameters on mitigating instabilities caused by the instability-inducing parameters $I$ as identified in Table~\ref{tab:GFM_param}, for static and dynamic lines, respectively.
\begin{table}[ht!]
    \vspace{-3ex}
    \begin{center}
  \setlength\tabcolsep{0.3pt}
  \setlength\extrarowheight{2pt}
  \caption{\vspace{0.8ex} \hspace{2ex} Stability Margin Sensitivity $\big(\Delta^{I|C}_{\lambda_0}\big)$ of GFM with static line.}
  \vspace{-3ex}
    \begin{tabular}{|r||*{12}{c|}}
      \Xhline{2\arrayrulewidth} \vspace{-0.7ex}
      \diagbox[outerleftsep=-3pt,outerrightsep=-1.8pt]{\scriptsize{$I\vspace{1ex}$}}{\scriptsize{$C\hspace{0.35ex}$}} & \scriptsize{$K^\text{P}_\text{VC}$} & \scriptsize{$K^\text{I}_\text{VC}$} & \scriptsize{$K^\text{F}_\text{VC}$} & \scriptsize{$K^\text{P}_\text{CC}$} & \scriptsize{$K^\text{I}_\text{CC}$} & \scriptsize{$K^\text{F}_\text{CC}$} & \scriptsize{$K_\text{P}$} &  \scriptsize{$K_\text{Q}$} & \scriptsize{$V_0$} & \scriptsize{$\omega_0$} & \scriptsize{$p^{*}$} & \scriptsize{$q^{*}$}\\
      \hline
      \scriptsize{$V_0$} & 0 & 0 & \cac{-7.311} & 0 & -0.874 & -2.154 & 0 & 0.064 & - & 0 & 2.479 & -0.01 \\
      \hline
      \scriptsize{$p^{*}$} & 0 & 0 & \cac{3.141} & 0 & 0.467 & 1.154 & 0 & -0.408 & 0.643 & 0 & - & 0.006 \\
      \hline
      \scriptsize{$L$} & 0 & 0 & \cac{1.76} & 0 & 0.465 & 1.154 & 0 & -0.014 & 0.043 & 0 & -1.036 & 0.0004 \\
      \hline
      \scriptsize{$R$} & 0 & 0 & \cac{0.775} & 0 & 0.082 & 0.203 & 0 & 0.342 & 0.501 & 0.014 & -0.347 & 0.005 \\
      \Xhline{2\arrayrulewidth} 
    \end{tabular}
    \label{tab:GFM_infl_S}
    \end{center}
\end{table}
\bac{For example, if instability arises due to changes in active power setpoint $p^{*}$, the most effective control action would be to adjust $K^\text{F}_\text{VC}$, as it has the largest sensitivity among all control parameters.}
\begin{table}[ht!]
    \vspace{-3ex}
    \begin{center}
  \setlength\tabcolsep{0.3pt}
  \setlength\extrarowheight{2pt}
  \caption{\vspace{0.8ex} \hspace{0ex} Stability Margin Sensitivity $\big(\Delta^{I|C}_{\lambda_0}\big)$ of GFM with dynamic line.}
  \vspace{-3ex}
    \begin{tabular}{|r||*{12}{c|}}
      \Xhline{2\arrayrulewidth} \vspace{-0.7ex}
      \diagbox[outerleftsep=-3pt,outerrightsep=-1.8pt]{\scriptsize{$I\vspace{1ex}$}}{\scriptsize{$C\hspace{0.35ex}$}} & \scriptsize{$K^\text{P}_\text{VC}$} & \scriptsize{$K^\text{I}_\text{VC}$} & \scriptsize{$K^\text{F}_\text{VC}$} & \scriptsize{$K^\text{P}_\text{CC}$} & \scriptsize{$K^\text{I}_\text{CC}$} & \scriptsize{$K^\text{F}_\text{CC}$} & \scriptsize{$K_\text{P}$} &  \scriptsize{$K_\text{Q}$} & \scriptsize{$V_0$} & \scriptsize{$\omega_0$} & \scriptsize{$p^{*}$} & \scriptsize{$q^{*}$}\\
      \hline
      \scriptsize{$V_0$} & 0 & 0 & \cac{-5.404} & -0.001 & -0.622 & -1.533 & 0 & 0.064 & - & 2.086 & 1.947 & -0.01 \\
      \hline
      \scriptsize{$p^{*}$} & 0 & 0 & \cac{2.823} & 0 & 0.393 & 0.972 & 0 &  -0.518 & 0.816 & -1.676 & - & 0.008\\
      \hline
      \scriptsize{$L$} & 0 & 0 & \cac{1.648} & 0 & 0.416 & 1.033 & 0 & -0.052 & 0.16 & -1.15 & -1.038 & 0.002 \\
      \hline
      \scriptsize{$R$} & 0 & 0 & \cac{0.942} & 0.028 & 0.114 & 0.282 & 0 & 0.475 & 0.697 & -0.058 & -0.275 & 0.007 \\
      \Xhline{2\arrayrulewidth} 
    \end{tabular}
    \label{tab:GFM_infl_D}
    \end{center}
\end{table}

Voltage instability caused by high active loading generally requires increased reactive power support for mitigation \cite{tamimi2010effect}. While for the GFL case, $q^{*}$ was identified as the most effective parameter for counteracting such instabilities, Tables \ref{tab:GFM_infl_S} and \ref{tab:GFM_infl_D} reveal that for \bac{GFM systems, the feedforward gain of the voltage controller $K^\text{F}_\text{VC}$ is the most effective parameter} in mitigating saddle-node bifurcations caused by any parameter. \bac{Interestingly, our previous study \cite{chatterjee2024effects} on Hopf bifurcations for GFM inverters identified the same parameter, $K^\text{F}_\text{VC}$, as most effective in mitigating oscillatory instability.}

\subsubsection{Example on Stability Margin Estimation}
Consider the active power setpoint $p^{*}$ as the instability-inducing parameter and $q^{*}$ as the control parameter. Tables \ref{GFM_stat} and \ref{GFM_dyn} compare the true margin with the estimated margin for systems with static and dynamic transmission lines, respectively. \bac{Both tables show that the normal vector method provides an effective estimate of the stability margins, emphasizing the insignificance of line dynamics.}

\begin{table}[ht!]
\vspace{-2ex}
\begin{center}
\caption{Effects of parameter variation \bac{(in per units)} on stability margin for GFM: Static line.}
\vspace{-1ex}
\begin{tabular}{cccc}
\Xhline{2\arrayrulewidth}
\multirow{2}{*}{$q^{*}\ \bac{\uparrow}$} & Estimated Margin & True Margin & \multirow{2}{*}{Error} \\
& in $p^{*}$ $\bac{\uparrow}$ & in $p^{*}$ & \\
\hline
        0.5 (Nominal) & - & 0.495078 & - \\
        0.6 & 0.501478 & 0.49683 & 4.64e-03 \\
         0.7 & 0.50788 & 0.49859 & 9.28e-03 \\
         0.9 & 0.52068 & 0.502105 & 1.86e-02 \\
          \Xhline{2\arrayrulewidth}
\end{tabular}
\label{GFM_stat}
\end{center}
\end{table}

\begin{table}[ht!]
\vspace{-7ex}
\begin{center}
\caption{Effects of parameter variation \bac{(in per units)} on stability margin for GFM: Dynamic line.}
\vspace{-1ex}
\begin{tabular}{cccc}
\Xhline{2\arrayrulewidth}
\multirow{2}{*}{$q^{*}\ \bac{\uparrow}$} & Estimated Margin & True Margin & \multirow{2}{*}{Error} \\
& in $p^{*}$ $\bac{\uparrow}$ & in $p^{*}$ & \\
\hline
        0.5 (Nominal) & - & 0.495078 & - \\
        0.6 & 0.503278 & 0.49683 & 6.44e-03 \\
         0.7 & 0.511478 & 0.49859 & 1.29e-02 \\
         0.9 & 0.527878 & 0.502107 & 2.58e-02 \\
          \Xhline{2\arrayrulewidth}
\end{tabular}
\label{GFM_dyn}
\end{center}
\vspace{-5ex}
\end{table}

\section{Conclusion}   
This paper focuses on voltage stability in inverter-based resource (IBR) systems and investigates the impacts of parameters and line dynamics on fold/saddle-node bifurcations. 
By deriving an analytical expression for the stability margin sensitivity using the normal vector to the bifurcation hypersurface, the most effective control parameters for mitigating fold/saddle-node bifurcations are identified for two common grid-following (GFL) and grid-forming (GFM) inverters. \bac{In GFL systems, the reactive loading setpoint and the current controller's feedforward gain were identified as the most effective parameters for improving voltage stability margins under high active and reactive loadings, respectively. On the other hand, the voltage controller's feedforward gain plays a dominant role in stabilizing GFM systems.} Furthermore, the analysis confirmed, both theoretically and numerically, that line dynamics have no impact on fold/saddle-node bifurcations in the considered GFM and GFL systems. This insight suggests the use of simpler static line models for computing stability margins and parameter sensitivities while studying (short-term) voltage stability. Future research will extend the approach to large-scale IBR-dominated power systems, where justifying the use of static line models over dynamic ones becomes important due to the significant complexity reduction.

\section*{Appendix}
\subsection{GFL Inverter Model}\label{app:detailed_model_gfl}
\begin{figure}[ht!]
    \vspace{-3ex}
    \hspace{-3ex} \centerline{\includegraphics[scale=0.8]{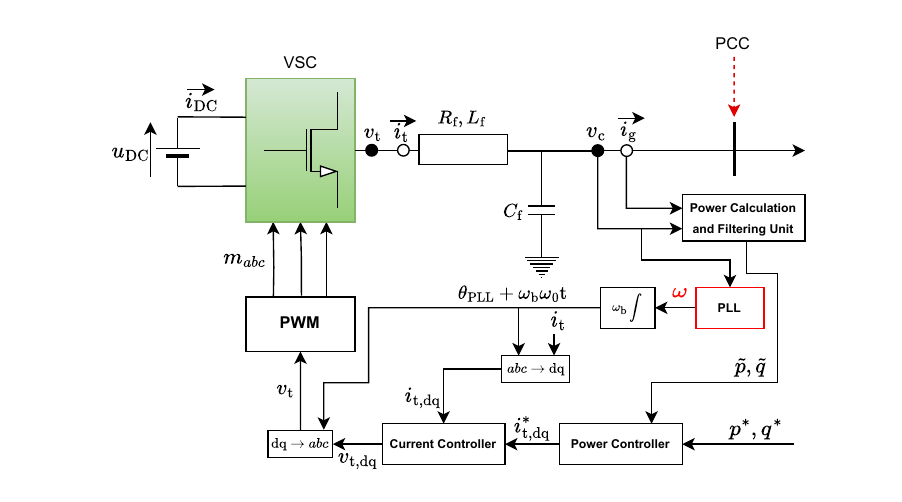}}
    \vspace{-2ex}
    \caption{Schematic of GFL inverter.}
    \label{fig:setup_gfl} 
\end{figure}

\begin{figure}[ht!]
    \vspace{-2ex} \hspace{11ex}\centerline{\includegraphics[scale=0.95]{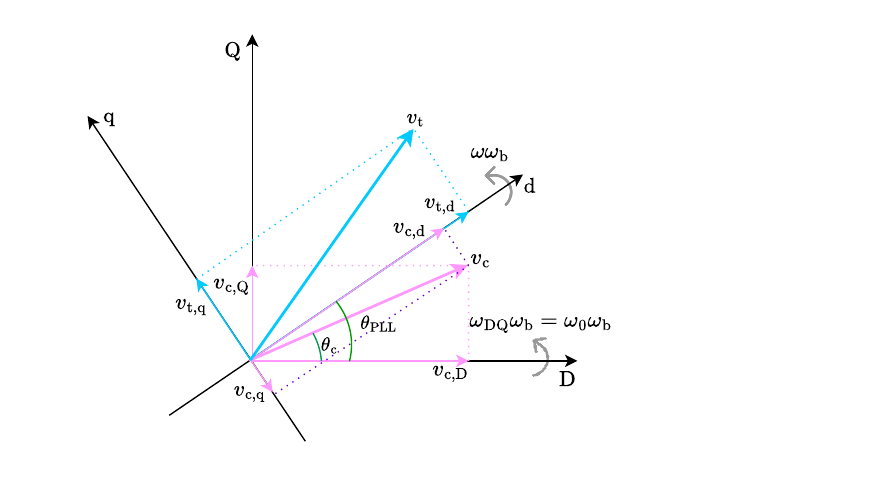}}
    \vspace{-8ex}
    \caption{Phasor diagram and reference frames of the GFL inverter.}
    \label{fig:gfl_ph} 
\end{figure}

\begin{figure*}[ht!]
    \centerline{\includegraphics[scale=0.72]{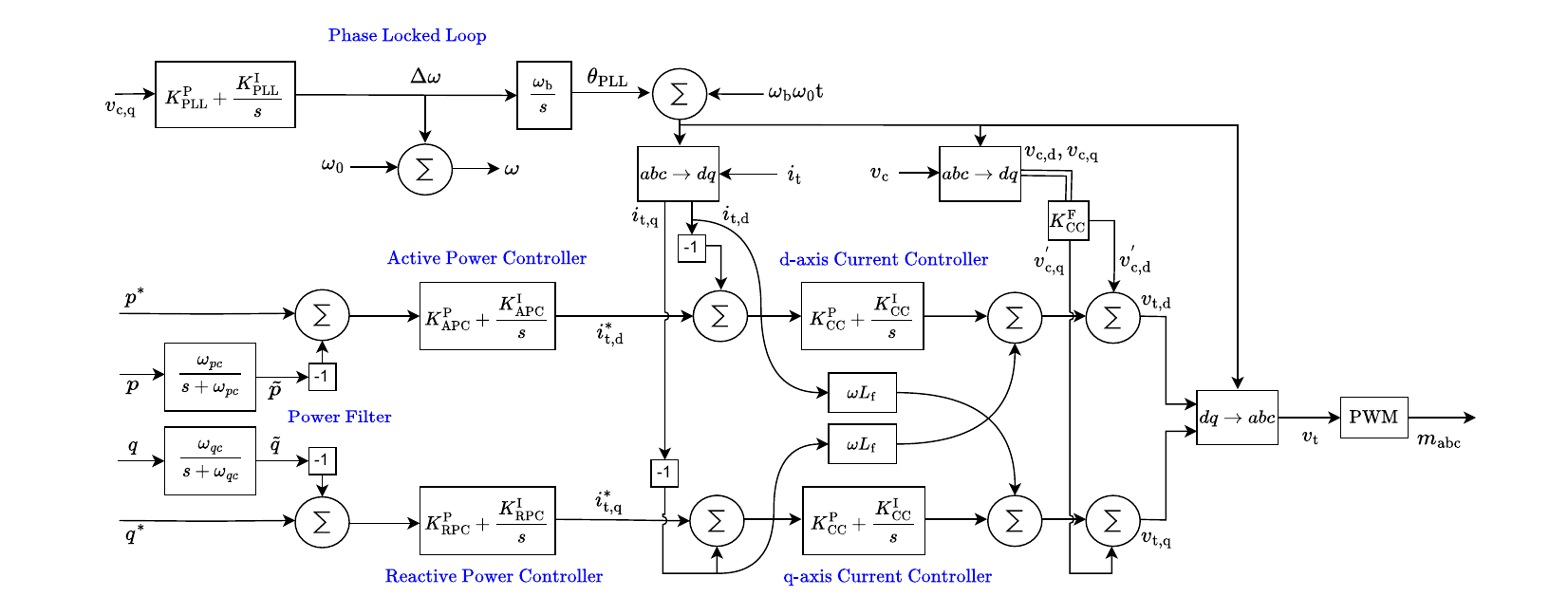}}
    \caption{Detailed control blocks of the GFL inverter.}
    \label{fig:control_gfl} 
\end{figure*}

We present the model for a commonly used GFL inverter as shown in Fig. \ref{fig:setup_gfl}.
The GFL inverter consists of a voltage source converter (VSC) with an output RLC filter, as described by equations \eqref{eqn:sl5}-\eqref{eqn:sl2}. Its control system features a hierarchical structure with an outer power control loop (active \eqref{eqn:sl9} and reactive \eqref{eqn:sl10} power controllers), which use filtered power measurements from \eqref{eqn:sl11}-\eqref{eqn:sl12} derived from \eqref{eqn:al7}-\eqref{eqn:al8}, and an inner current-control loop. The outer loop generates reference signals \eqref{eqn:al12}-\eqref{eqn:al13}, which are fed into the inner current-control loop as described by \eqref{eqn:sl3}-\eqref{eqn:sl4} and \eqref{eqn:al3}-\eqref{eqn:al4}. DC power balance is managed by \eqref{eqn:al14}. All control loops are implemented in a local dq-reference frame, with a phase-locked loop (PLL) \eqref{eqn:sl7} used to generate the angle \eqref{eqn:sl8} and frequency \eqref{eqn:al9}-\eqref{eqn:al9.5} for converting three-phase $abc$ signals to dq quantities. The transformation from local to global DQ-frame is detailed in equations \eqref{eqn:al10}-\eqref{eqn:al2}, and illustrated in Fig. \ref{fig:gfl_ph}. The inverter's detailed control loops are given in Fig. \ref{fig:control_gfl}. All equations and variables are in per unit unless noted otherwise.

\subsubsection{Nomenclature}
The variables and parameters are indexed below.\\ 
\texttt{Dynamic states (12):}\\
\begin{small}
$x_{1}\ :\ \tilde{p} \rightarrow$ Filtered active power output at PCC\\
$x_{2}\ :\ \tilde{q} \rightarrow$ Filtered reactive power output at PCC\\
$x_3\ :\ \gamma_\text{PLL} \rightarrow$ PLL integrator state\\
$x_4\ :\ \theta_\text{PLL} \rightarrow$ Angle of the local dq-frame w.r.t nominal frame\\
$x_5\ :\ \phi_\text{d} \rightarrow$ Active power controller state\\
$x_6\ :\ \phi_\text{q} \rightarrow$ Reactive power controller state\\
$x_7\ :\ \gamma_\text{d} \rightarrow$ d-axis current controller state\\
$x_8\ :\ \gamma_\text{q} \rightarrow$  q-axis current controller state\\
$x_9\ :\ v_\text{c,d} \rightarrow$ d-axis filter output voltage $v_\text{c}$ at PCC\\
$x_{10}\ :\ v_\text{c,q} \rightarrow$ q-axis filter output voltage $v_\text{c}$ at PCC\\
$x_{11}\ :\ i_\text{t,d} \rightarrow$ d-axis inverter output current $i_\text{t}$\\
$x_{12}\ :\ i_\text{t,q} \rightarrow$ q-axis inverter output current $i_\text{t}$
\end{small}
\newline\newline
\texttt{Algebraic states (16):}\\
\begin{small}
$y_1\ :\ \omega \rightarrow$ Angular frequency of inverter’s local dq-frame per PLL\\
$y_2\ :\ i^{*}_\text{t,d} \rightarrow$ d-axis current setpoint $i^{*}_\text{t}$\\
$y_3\ :\ i^{*}_\text{t,q} \rightarrow$  q-axis current setpoint $i^{*}_\text{t}$\\
$y_4\ :\ i_\text{g,d} \rightarrow$ d-axis filter output current $i_\text{g}$ at PCC\\
$y_5\ :\ i_\text{g,q} \rightarrow$ q-axis filter output current $i_\text{g}$ at PCC\\
$y_6\ :\ i_\text{g,D} \rightarrow$ D-axis filter output current $i_\text{g}$ at PCC\\
$y_7\ :\ i_\text{g,Q} \rightarrow$ Q-axis filter output current $i_\text{g}$ at PCC\\
$y_8\ :\ v_\text{c,D} \rightarrow$ D-axis filter output voltage $v_\text{c}$ at PCC\\
$y_9\ :\ v_\text{c,Q} \rightarrow$ Q-axis filter output voltage $v_\text{c}$ at PCC\\
$y_{10}\ :\ p \rightarrow$ Active power output at PCC \\
$y_{11}\ :\ q \rightarrow$ Reactive power output at PCC \\
$y_{12}\ :\ v_\text{t,d} \rightarrow$ d-axis inverter output voltage $v_\text{t}$\\
$y_{13}\ :\ v_\text{t,q} \rightarrow$ q-axis inverter output voltage $v_\text{t}$\\
$y_{14}\ :\ \Delta \omega \rightarrow$ Angular frequency error of dq-frame w.r.t nominal\\
$y_{15}\ :\ u_\text{DC} \rightarrow$ DC-link voltage\\
$y_{16}\ :\ i_\text{DC} \rightarrow$ DC-link current
\end{small}
\newline\newline
\texttt{Parameters (18):}\\
\begin{small}
$p_1\ :\ p^{*} \rightarrow$ Active power setpoint of inverter\\
$p_2\ :\ q^{*} \rightarrow$ Reactive power setpoint of inverter\\
$p_3\ :\ \omega_0 \rightarrow$ Nominal grid frequency \\
$p_4\ :\ \omega_{pc} \rightarrow$ Active power filter 3dB cut-off frequency \\
$p_5\ :\ \omega_{qc} \rightarrow$ Reactive power filter 3dB cut-off frequency \\
$p_6\ :\ \omega_\text{b} \rightarrow$ Base value of angular frequency in rad/s\\
$p_7\ :\ K_\text{PLL}^\text{P} \rightarrow$ Proportional gain of PLL \\
$p_8\ :\ K_\text{PLL}^\text{I} \rightarrow$ Integral gain of PLL \\
$p_9\ :\ K_\text{APC}^\text{P} \rightarrow$  Proportional gain of active power controller \\
$p_{10}\ :\ K_\text{APC}^\text{I} \rightarrow$  Integral gain of active power controller \\
$p_{11}\ :\ K_\text{RPC}^\text{P} \rightarrow$ Proportional gain of reactive power controller \\
$p_{12}\ :\ K_\text{RPC}^\text{I} \rightarrow$ Integral gain of reactive power controller \\
$p_{13}\ :\ K_\text{CC}^\text{P} \rightarrow$  Proportional gain of current controller \\
$p_{14}\ :\ K_\text{CC}^\text{I} \rightarrow$  Integral gain of current controller \\
$p_{15}\ :\ K_\text{CC}^\text{F} \rightarrow$  Feed-forward gain of current controller \\
$p_{16}\ :\ C_\text{f} \rightarrow$ Capacitance of output filter \\
$p_{17}\ :\ L_\text{f} \rightarrow$ Inductance of output filter \\
$p_{18}\ :\ R_\text{f} \rightarrow$ Resistance of output filter 
\end{small}

\subsubsection{Differential-Algebraic Equations (DAEs)}
\texttt{Differential equations:}
\begin{flalign}
        \dot{\tilde{p}} &\ =\ -\omega_{pc} (\tilde{p} - p) \label{eqn:sl11}
        \\
        \dot{\tilde{q}} &\ =\ -\omega_{qc} (\tilde{q} - q) \label{eqn:sl12}
        \\
        \dot{\gamma_\text{PLL}} &\ =\ v_\text{c,q} \label{eqn:sl7}
        \\
        \dot{\theta_\text{PLL}} &\ =\ \omega_\text{b} \Delta \omega \label{eqn:sl8}
        \\
        \dot{\phi}_\text{d} &\ =\ p^{*} - \tilde{p} \label{eqn:sl9}
        \\
        \dot{\phi}_\text{q} &\ =\ q^{*} - \tilde{q} \label{eqn:sl10}
        \\
        \dot{\gamma_\text{d}} &\ =\ i^{*}_\text{t,d} - i_\text{t,d} \label{eqn:sl3}
        \\
        \dot{\gamma_\text{q}} &\ =\ i^{*}_\text{t,q} - i_\text{t,q} \label{eqn:sl4}
        \\
        \dot{v}_\text{c,d} &\ =\ \omega_\text{b} \omega v_\text{c,q} + \frac{\omega_\text{b}}{C_\text{f}}(i_\text{t,d} - i_\text{g,d}) \label{eqn:sl5}
        \\
        \dot{v}_\text{c,q} &\ =\ -\omega_\text{b} \omega v_\text{c,d} + \frac{\omega_\text{b}}{C_\text{f}}(i_\text{t,q} - i_\text{g,q}) \label{eqn:sl6}
        \\
        \dot{i_\text{t,d}} &\ =\ \omega_\text{b} \omega i_\text{t,q} + \frac{\omega_\text{b}}{L_\text{f}}(v_\text{t,d} - v_\text{c,d}) - \frac{R_\text{f}}{L_\text{f}}\omega_\text{b} i_\text{t,d}
        \label{eqn:sl1} &&
        \\
        \dot{i_\text{t,q}} &\ =\ - \omega_\text{b} \omega i_\text{t,d} + \frac{\omega_\text{b}}{L_\text{f}}(v_\text{t,q} - v_\text{c,q})  - \frac{R_\text{f}}{L_\text{f}}\omega_\text{b} i_\text{t,q}
        \label{eqn:sl2}
\end{flalign}
\noindent 
\texttt{Algebraic equations:}
\begin{align}
    &\ \omega - \omega_0 - \Delta \omega = 0 \label{eqn:al9}
    \\&\
    \Delta \omega - K^\text{P}_\text{PLL}v_\text{c,q} - K^\text{I}_\text{PLL}\gamma_\text{PLL} = 0 \label{eqn:al9.5}
    \\&\
    i_\text{t,d}^{*} - K^\text{P}_\text{APC}(p^{*}-\tilde{p}) - K^\text{I}_\text{APC}\phi_\text{d} = 0 \label{eqn:al12}
    \\&\
    i_\text{t,q}^{*} - K^\text{P}_\text{RPC}(q^{*}-\tilde{q}) - K^\text{I}_\text{RPC}\phi_\text{q} = 0 \label{eqn:al13}
    \\&\
    v_\text{t,d} - K^\text{P}_\text{CC}(i^{*}_\text{t,d}\!-\!i_\text{t,d}) - K^\text{I}_\text{CC} \gamma_\text{d} - K^\text{F}_\text{CC}v_\text{c,d} + \omega L_\text{f} i_\text{t,q} = 0 \label{eqn:al3}
    \\&\
    v_\text{t,q} - K^\text{P}_\text{CC}(i^{*}_\text{t,q}\!-\!i_\text{t,q}) - K^\text{I}_\text{CC} \gamma_\text{q} - K^\text{F}_\text{CC}v_\text{c,q} - \omega L_\text{f} i_\text{t,d} = 0 \label{eqn:al4}
    \\&\ 
    i_\text{g,d} - i_\text{g,D}\cos(\theta_\text{PLL}) - i_\text{g,Q}\sin(\theta_\text{PLL}) = 0 \label{eqn:al10}
    \\&\
    i_\text{g,q} + i_\text{g,D}\sin(\theta_\text{PLL}) - i_\text{g,Q}\cos(\theta_\text{PLL}) = 0 \label{eqn:al11}    
    \\&\ 
    v_\text{c,d} - v_\text{c,D}\cos(\theta_\text{PLL}) - v_\text{c,Q}\sin(\theta_\text{PLL}) = 0 \label{eqn:al1}
    \\&\
    v_\text{c,q} + v_\text{c,D}\sin(\theta_\text{PLL}) - v_\text{c,Q}\cos(\theta_\text{PLL}) = 0 \label{eqn:al2}
    \\&\
    p - v_\text{c,d}i_\text{g,d} - v_\text{c,q}i_\text{g,q} = 0 \label{eqn:al7}
    \\&\ 
    q - v_\text{c,q}i_\text{g,d} + v_\text{c,d}i_\text{g,q} = 0 \label{eqn:al8}
    \\&\
    u_\text{DC}i_\text{DC} - v_\text{t,d}i_\text{t,d} - v_\text{t,q}i_\text{t,q} = 0 \label{eqn:al14}
\end{align}

\bibliographystyle{ieeetr}
\bibliography{Bibliography.bib}
\balance
\end{document}